\documentclass[aps,prb,preprint,groupedaddress]{revtex4}

\usepackage{graphicx}
\usepackage{dcolumn}
\usepackage{epsfig}
\usepackage{bm}

\newcommand{\be}{\begin{equation}}
\newcommand{\ee}{\end{equation}}
\newcommand{\bea}{\begin{eqnarray}}
\newcommand{\eea}{\end{eqnarray}}

\begin{document}


\title{Series Expansions for Variable Electron Density}


\author{Weihong Zheng}
\email[]{w.zheng@unsw.edu.au}
\homepage[]{http://www.phys.unsw.edu.au/~zwh}
\affiliation{School of Physics,
The University of New South Wales,
Sydney, NSW 2052, Australia.}

\author{C.J. Hamer}
\email[]{c.hamer@unsw.edu.au}
\affiliation{School of Physics,
The University of New South Wales,
Sydney, NSW 2052, Australia.}

\author{J. Oitmaa}
\email[]{j.oitmaa@unsw.edu.au}
\affiliation{School of Physics,
The University of New South Wales,
Sydney, NSW 2052, Australia.}

\author{R.R.P. Singh}
\email[]{singh@raman.ucdavis.edu}
\affiliation{Department of Physics, University of California, Davis,
CA95616, USA}


\date{\today}

\begin{abstract}
We have developed a series expansion method 
for calculating the zero-temperature properties of lattice electron models for
variable electron density, i.e. for finite doping away from the half-filled case.
This is done by 
introducing particle fluctuation terms in both the
 unperturbed Hamiltonian and perturbation.
The method is demonstrated by application to the 
 2-chain $t-J$ ladder, where we provide comparison with previous work as 
 well as obtaining a number of new results.
\end{abstract}


\maketitle

\section{\label{sec:intro}INTRODUCTION}
The last 15 years have seen an explosion of interest in the study of
lattice models of strongly correlated electrons. While, in large part,
this has been driven by the field of high $T_c$ superconductivity,
there are other classes of materials whose properties are also determined
by strong electron correlations, such as Kondo insulators, heavy
fermion systems, and organic conductors.

Models for such systems usually start from the well known Hubbard
and/or $t-J$ models, or generalizations of these. Exact solutions
for these models are known only in 1-dimension. In higher
dimensions, and indeed in 1-dimension for non-solvable cases, a variety
of analytic and numerical techniques have been used: variational
methods, exact diagonalizations for small systems, quantum Monte
Carlo approaches, the density matrix renormalization group (DMRG),
and perturbative ``linked-cluster" series
expansions. Yet it is well-known that strong-correlated electron
models cause difficulties for most numerical techniques.
Quantum Monte Carlo simulations are subject to the infamous
``minus sign" problem, while exact diagonalization and DMRG calculations
suffer from finite-size effects. It is therefore important to develop new
techniques which may help to shed light on these problem.

In this paper we focus on series expansions at zero temperature\cite{gel00}.
This approach has been used to study both the Hubbard model\cite{shi95,zhe01}
and the $t\!-\!J$ model\cite{cjh98,oit99}. In the past this method has been
restricted to states with one electron per site (half-filling) and
to one and two-hole excitations away from half-filling.
While this region of the phase diagram is itself of considerable
interest, many of the more interesting aspects of strongly correlated systems
occur for finite density of holes (``finite doping").

We have explored a number of ways in which the series method might be adapted
to variable electron density and have been able to obtain consistent
results for one model of considerable current interest,
the $t-J$ ladder\cite{oit99,poi95,mul98}. The aim of this paper is thus
twofold: to demonstrate the method, and to present new results for the
$t\!-\!J$ ladder away from half-filling.

For completeness we summarize what is known for the $t-J$ ladder,
both at half-filling and for finite doping.
A clear discussion has been given by Troyer, Tsunetsugu and  Rice\cite{tro96}.
At half-filling, and in the limit of strong
interchain coupling, the ground state consists of spin singlet dimers  on
each rungs. The lowest spin excitation consists of a triplet 
excitation on one rung, propagating via the coupling
between rungs.
There are also quasiparticle excitations ($S=\frac{1}{2}$),
corresponding in the limit to a hole excitation on a single rung
$-$these carry both spin and charge, and separation of spin and
charge does not occur in this model.

The lowest 2-hole excitation consists of a 
singlet hole pair on one rung, which develop into a
band of states by propagation along the ladder.
The zero-momentum state in this band is then the ground state in the 2-hole sector,
and then the other states correspond to a gapless band of particle-hole charge
excitations relative to the ground state.
Correspondingly, the system is said to be in a C1S0) phase (1 gapless charge
mode, 0 gapless spin modes).

A similar picture appear to hold as one moves away from the strong interchain
coupling limit, even in the isotropic case which has been the object of
most studies. As one moves away from half-filling, however, one finds that
the lowest triplet excitation corresponds not to the simple triplet excitation of a singlet dimer,
but to a particle-hole pair excitation on different rungs.
Thus the triplet gap evolves discontinuously away from half-filling.\cite{tro96}

%
Poilblanc, Scalapino and   Hanke \cite{poi95} and M\"uller and Rice\cite{mul98}
have presented a plausible phase diagram in the $n$ versus $J/t$
plane (for the isotropic case). For $J/t\lesssim 2$ the system is
in a C1S0 phase (1 gapless charge mode) for low to moderate doping
and crosses to a C1S1 phase (i.e. the spin-gap vanishes) for
higher doping. For $J/t\gtrsim 2$ phase separation is predicted to
occur. For small $J/t$ and low doping M\"uller and
Rice\cite{mul98} also find ferromagnetic (Nagaoka) and C2S2 phases.

The paper is arranged as follows: in Sect. II we describe our series expansion method;
Section III and IV present results, for a case with strong interaction coupling and for
the isotropic case; Section IV gives a summary and discussion.


\section{\label{sec:method}Series Expansions for Variable Electron Density}

The Hamiltonian of the $t-J$ ladder, in usual notation, is
\bea
H = && J \sum_{i,a} ( {\bf S}_{i,a} \cdot {\bf S}_{i+1,a}
   -{1\over 4} n_{i,a} n_{i+1,a} )
  + J_{\perp} \sum_{i} ( {\bf S}_{i,1} \cdot {\bf S}_{i,2}
   -{1\over 4} n_{i,1} n_{i,2} ) \nonumber \\
 && - t \sum_{i,a,\sigma} P (c^{\dag}_{i,a,\sigma} c_{i+1,a,\sigma}
        + H.c. ) P  - t_{\perp} \sum_{i,\sigma} P (c^{\dag}_{i,1,\sigma} c_{i,2,\sigma}
        + H.c. ) P  \label{eqH}
\eea
where $i$ labels sites along each chain, $\sigma$ ($=\uparrow$ or
$\downarrow$) and $a$ (=1,2) are spin and leg indices,
 $P$ is a projection
operator which excludes doubly occupied sites.
$J$, $t$ are exchange and hopping parameters on each chain,
while $J_{\perp}$, $t_{\perp}$ are coupling parameters
between the two chains, i.e. on the rungs of the ladder.

In our previous work, and also in the present work, we employ a
``rung basis", in which the 2nd and 4th terms in (\ref{eqH})
form the unperturbed Hamiltonian, and the remaining terms are
treated perturbatively. The ground state of the unperturbed Hamiltonian,
at half-filling, is then a direct product of spin-singlet states on each
rung. Spin excitations, at half-filling, consist of a spin-triplet
on one rung, which propagates coherently along the ladder. One and
two-hole charge excitations are created by removing one or two electrons
from a rung state and allowing these to propagate, via the $t$-hopping
term.

At finite doping the situation is more complex, and it is impossible
to identify a suitable unperturbed ground state. However it is possible to
make progress by relaxing the constraint on particle number, by adding
a particle nonconserving term to the Hamiltonian, and by introducing
a chemical potential to control the electron density. A variety of particle nonconserving terms are
possible, but we have found the following form
$$
{h\over \sqrt{2}} \sum_{i,a,\sigma}  ( c^{\dag}_{i,1,\uparrow} c^{\dag}_{i,2,\downarrow}
- c^{\dag}_{i,1,\downarrow} c^{\dag}_{i,2,\uparrow} + H.c. ),
$$
to give best results. This term creates a spin-singlet on an empty rung or
destroys a spin-singlet state to create an empty rung.

Our Hamiltonian, then, is
\be
H= H_0 + x V
\ee
where
\bea
H_0 &=& J_{\perp} \sum_{i} ( {\bf S}_{i,1} \cdot {\bf S}_{i,2}
   -{1\over 4} n_{i,1} n_{i,2} )
   - t_{\perp} \sum_{i,\sigma} P (c^{\dag}_{i,1,\sigma} c_{i,2,\sigma}
        + H.c. ) P \nonumber \\
&&   + {h\over \sqrt{2}} \sum_{i} P ( c^{\dag}_{i,1,\uparrow} c^{\dag}_{i,2,\downarrow}
- c^{\dag}_{i,1,\downarrow} c^{\dag}_{i,2,\uparrow} + H.c. ) P
-\mu \sum_i (n_{i,1} + n_{i,2} ) \nonumber \\
V &=&  J \sum_{i,a} ( {\bf S}_{i,a} \cdot {\bf S}_{i+1,a}
   -{1\over 4} n_{i,a} n_{i+1,a} )
   - t \sum_{i,a,\sigma} P (c^{\dag}_{i,a,\sigma} c_{i+1,a,\sigma}
        + H.c. ) P \nonumber \\
&& -  {h\over \sqrt{2}} \sum_{i} P ( c^{\dag}_{i,1,\uparrow} c^{\dag}_{i,2,\downarrow}
- c^{\dag}_{i,1,\downarrow} c^{\dag}_{i,2,\uparrow} + H.c. ) P
\eea
and $x$ is an expansion parameter.
The particle fluctuation term will mix sectors with different particle number, but
these terms cancel in the physical limit $x=1$, and the final results should be
independent of the field $h$. In practice $h$ can be adjusted to improve convergence
for the extrapolation $x\to 1$. The electron density is determined by the
chemical potential term, as usual.

The eigenstates of $H_0$ are direct products constructed
from the nine possible rung states, which are given in Table I.
For the range of parameters we use, the lowest energy rung state is
\be
\vert \chi \rangle =
(\kappa -J_{\perp} -2 \mu) \mid 00 \rangle -
  \sqrt{2} h (\mid \uparrow \downarrow \rangle - \mid \downarrow \uparrow \rangle )
\ee
where $\kappa$ is defined in Table 1. This is a spin-singlet state.

To compute the perturbation series we choose values for the free
parameters $t,h,\mu$, and derive expansions in powers of $x$ for the quantities
of interest. Series have been computed to order $x^{10}$ for the ground state
energy $E_0$, electron density $n$, and the dispersion relations for
spin-triplet excitations $\Delta_s (k)$ and one-hole bonding excitations
$\Delta_c (k)$. The calculations involve (trivial) 1-dimensional
clusters up to 11 rungs, and are limited to this order by computer
memory constraints. The series are then evaluated at $x=1$, corresponding to
the full Hamiltonian, by integrated differential approximants\cite{gut}. The series
are too numerous to be reproduced here, but are available on request.

\section{\label{sec:res}Results for the Doped \mbox{$t-J$} ladder (Strong Rungs)}

As our series expansion is about the rung limit, convergence will naturally be better
when the rung coupling terms are greater than coupling along the chains. We present
in this Section results for 
$J_{\perp}/J=4$.

\begin{figure}[ht] 
\vspace{0cm}
\centerline{\hbox{\psfig{figure=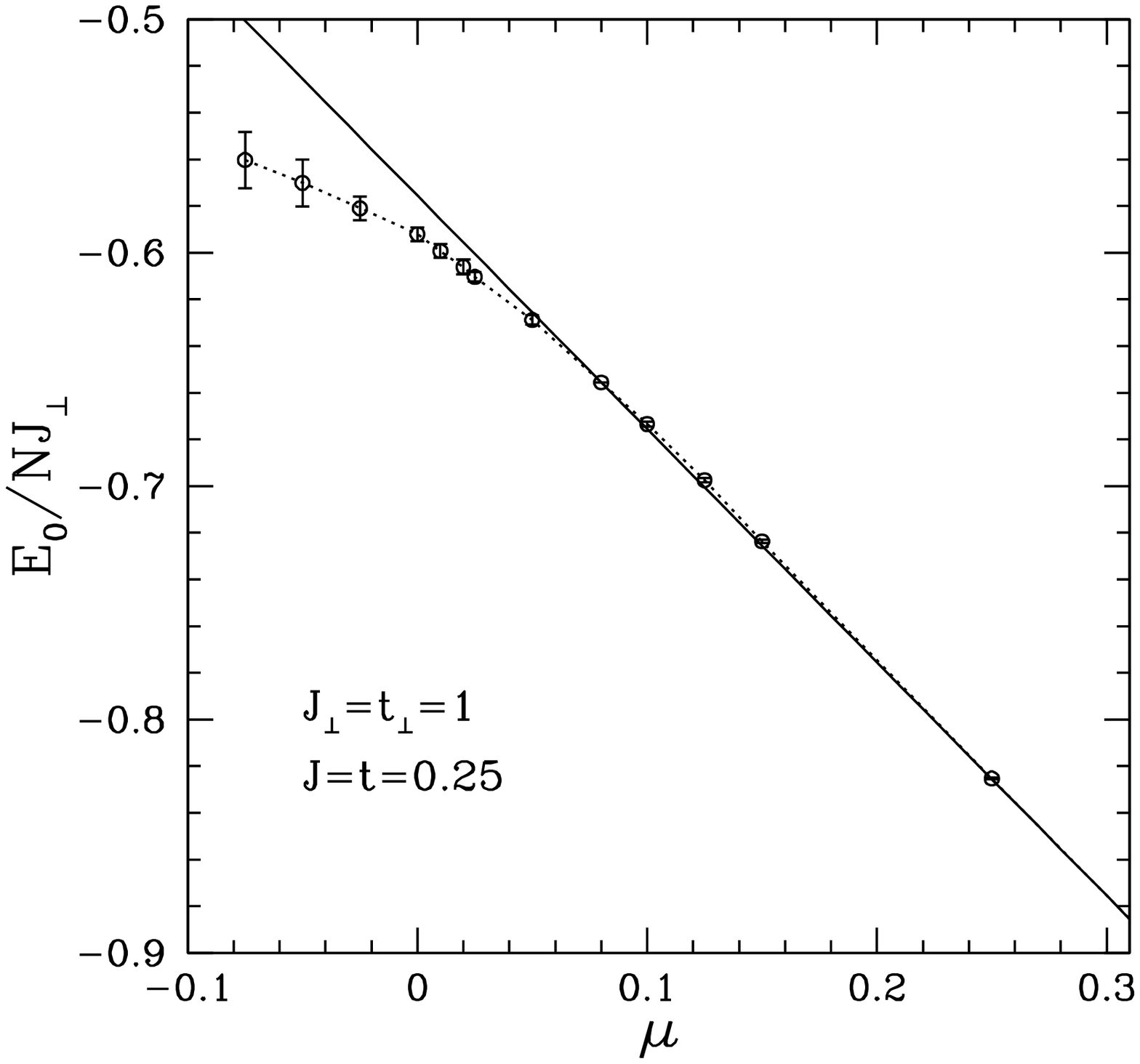,width=15cm}}}
\par
\caption{The ground state energy per site $E_0/NJ$
versus $\mu$ for $J_{\perp}=t_{\perp}=1$, $J=t=0.25$.
The solid line is the result at
half-filling\cite{zwh98}.
}
\label{fig_e0_mu_p25}
\end{figure}

\begin{figure}[ht] 
\vspace{0cm}
\centerline{\hbox{\psfig{figure=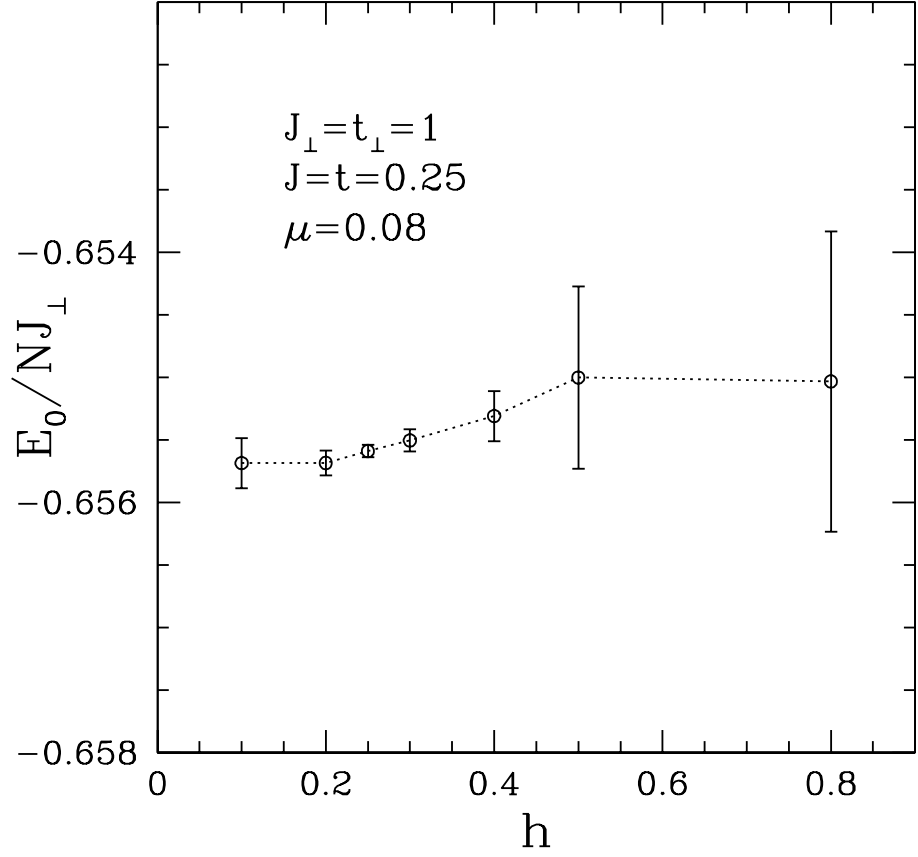,width=15cm}}}
\par
\caption{The ground state energy per site $E_0/NJ$
versus $h$ for $J_{\perp}=t_{\perp}=1$, $J=t=0.25$, and $\mu=0.08$.
}
\label{fig_e0_h}
\end{figure}

We first consider the ground state energy as a function of chemical potential $\mu$.
We have evaluated series, in powers of $x$ to order $x^{10}$, for
various fixed $h,\mu$. The leading terms are
\be
E_0/N = - {1\over 4} (J_{\perp} + 2 \mu + \kappa ) + x ({h^2\over \kappa} -
 {(J_{\perp} + 2 \mu + \kappa )^2 \over 16 \kappa^2 }J ) + O(x^2)
\ee
with $\kappa = \sqrt{4 h^2 + (J_{\perp}+2 \mu)^2 }$. Figure \ref{fig_e0_mu_p25} shows curves of
$E_0$ versus $\mu$ for $t=J=0.25$, $t_{\perp}=J_{\perp}=1$. The solid line in this
figure is the result for half-filling ($n=1$). For this case
the charge degrees of freedom are frozen out, and the system is
equivalent to a Heisenberg spin ladder, where the ground
state energy per site, for $J_{\perp}=4J$, is
$E_0/NJ_{\perp} = -0.38803708$\cite{zwh98}. Including the other terms
gives, at half-filling, $E_0/NJ =
 -0.38803708-3/16-\mu$, which is the solid line in the figure.
 For $\mu \gtrsim $ 0.05 our ground state energy is very close to that for
 half-filling. If we take a large value of $\mu$, $\mu=5$, to make sure
 the system is  half-filled at $x=1$,
 the ground state energy is estimated to be $E_0/NJ_{\perp} =-5.575537(1)$,
 which agrees with the results for the Heisenberg spin ladder to 6 digits.
 The results in Figure \ref{fig_e0_mu_p25} have been obtained by
 adjusting the field $h$ to obtain best convergence.
 An illustration of the sensitivity of this procedure is shown in
 Figure \ref{fig_e0_h}, where $E_0$ is plotted versus $h$, for
 $J_{\perp}=t_{\perp}=1$, $J=t=0.25$, and $\mu=0.08$. The size of 
 the error bars, which represent the spread
 among different approximants, is least for $h\sim 0.25$, which is the value used.
 $E_0$ itself is relatively insensitive to the value of $h$ chosen.

\begin{figure}[ht] 
\vspace{0cm}
\centerline{\hbox{\psfig{figure=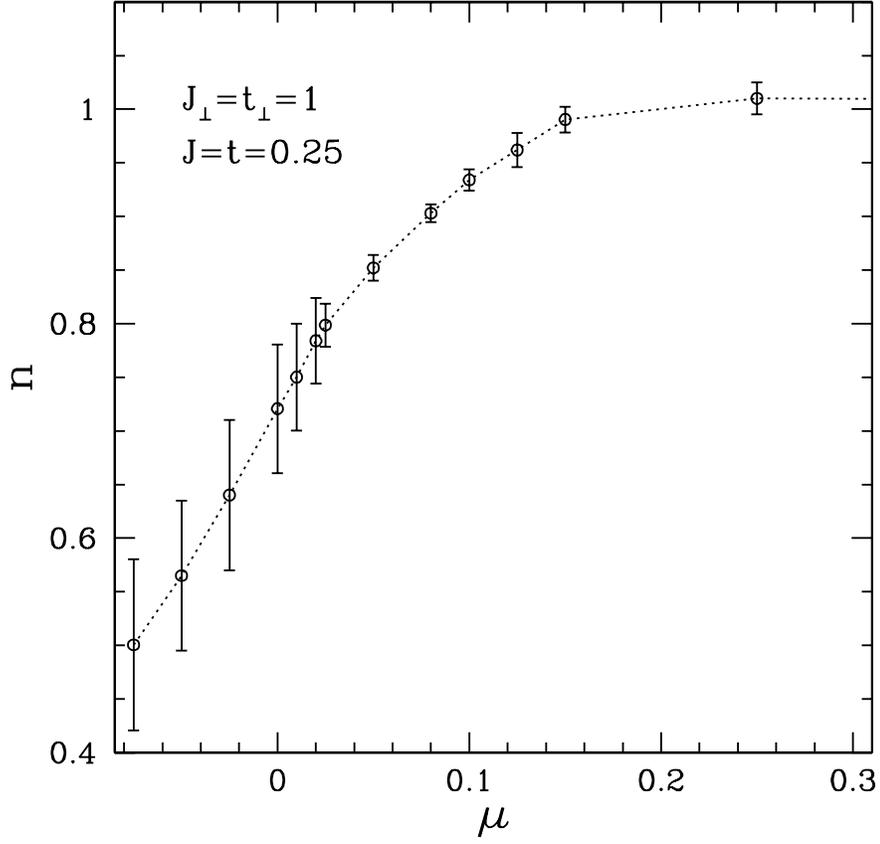,width=15cm}}}
\par
\caption{The electron density $n$ versus $\mu$ for $J_{\perp}=t_{\perp}=1$, $J=t=0.25$.
}
\label{fig_n_p25}
\end{figure}

The electron density is obtained via the standard relation
\be
n = - {\partial \over \partial \mu} {E_0 \over N}
\ee
from which we obtain series in $x$, which are again evaluated at $x=1$ via
approximants. Figure \ref{fig_n_p25} shows curves of electron density versus $\mu$
for the same parameters as Fig. \ref{fig_e0_mu_p25}. Half-filling corresponds to large $\mu$, as
expected. 
As $\mu$ decreases, the electron density begins to drop, and below
$\mu\simeq 0$ the error bars increase substantially, but the overall
trend is very clear.

\begin{figure}[ht] 
\vspace{0cm}
\centerline{\hbox{\psfig{figure=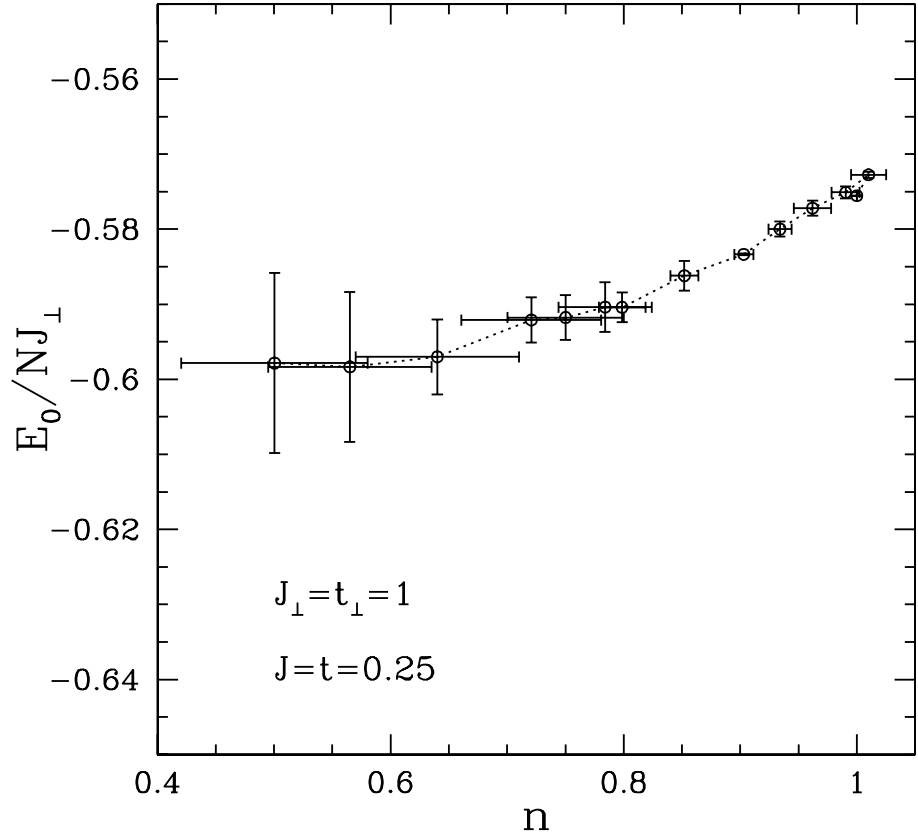,width=15cm}}}
\par
\caption{The ground state energy versus electron density $n$ for $J_{\perp}=t_{\perp}=1$, $J=t=0.25$.
}
\label{fig_e0_n_p25}
\end{figure}

We have used the data of Figures \ref{fig_e0_mu_p25} and \ref{fig_n_p25} 
 to compute the ground state energy as a function of $n$, and
this is shown in Figure \ref{fig_e0_n_p25}. In this figure
we have subtracted the chemical potential term $\mu n$ from the energy.
The ground state energy seems rather insensitive to the electron density 
from these parameters.

\begin{figure}[ht] 
\vspace{0cm}
\centerline{\hbox{\psfig{figure=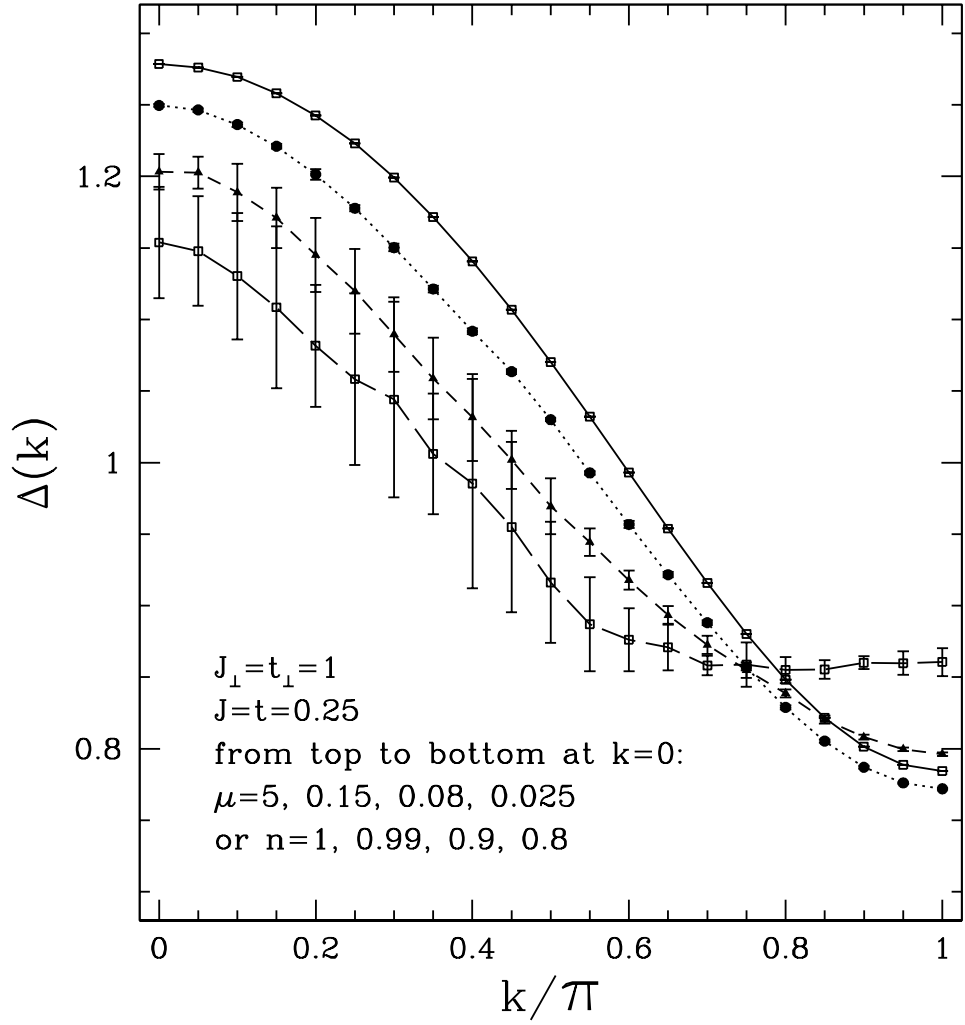,width=15cm}}}
\par
\caption{The triplet spin dispersion relation for $J_{\perp}=t_{\perp}=1$, $J=t=0.25$
and  $\mu=5,0.15, 0.08, 0.025$,  corresponding to $n=1,0.99,0.9,0.8$.
}
\label{fig_triplet_gap_p25}
\end{figure}

Next we consider the dispersion relation
$\Delta_s (k)$ for triplet spin excitations. At half-filling these excitations
correspond physically to a triplet excitation on a rung propagating coherently
along the ladder. At finite doping the situation is more complicated because
rung singlets contain an admixture of empty states, and singlet-triplet excitation
processes include pair creation terms. Nevertheless spin-triplet excitations
are well defined and their energy dispersion can be calculated via series
methods\cite{gel00,gel96}. Figure \ref{fig_triplet_gap_p25} shows the 
triplet spin dispersion curve $\Delta_s (k)$
for $J_{\perp}=t_{\perp}=1$, $J=t=0.25$, and $\mu = 5, 0.15, 0.08, 0.025$,
 which correspond to
$n=1.000000(1)$, 0.99(1), 0.90(1), 0.80(2), respectively. 
For $\mu=5$, the system is at half-filling, and we can reproduce the
dispersion curve for the Heisenberg spin ladder with $4J=J_{\perp}$\cite{zwh98}, up to 6 digits.
The figure shows that the triplet dispersion is very
sensitive to doping, with a significant change already for 1\%
doping ($\mu=0.15$). 
The minimum triplet gap remains at $k=\pi$ for low doping ($n\gtrsim 0.8$).
The size of the gap decreases for very low doping, but then begins to increase.
By $n=0.8$ the dispersion curve has become quite flat near the zone boundary, and
it appears that the minimum triplet energy shifts
to intermediate $k$ values for larger doping.

\section{\label{sec:res2}Results for the Doped \mbox{$t-J$} ladder (Isotropic case)}

Most previous studies of the $t-J$ ladder have been for the isotropic exchange
case $J=J_{\perp}$, $t=t_{\perp}$. Although our series here are less regular we are able to obtain
results which can be compared to previous work. We follow, more or less, the order of Section III.

\begin{figure}[ht] 
\vspace{0cm}
\centerline{\hbox{\psfig{figure=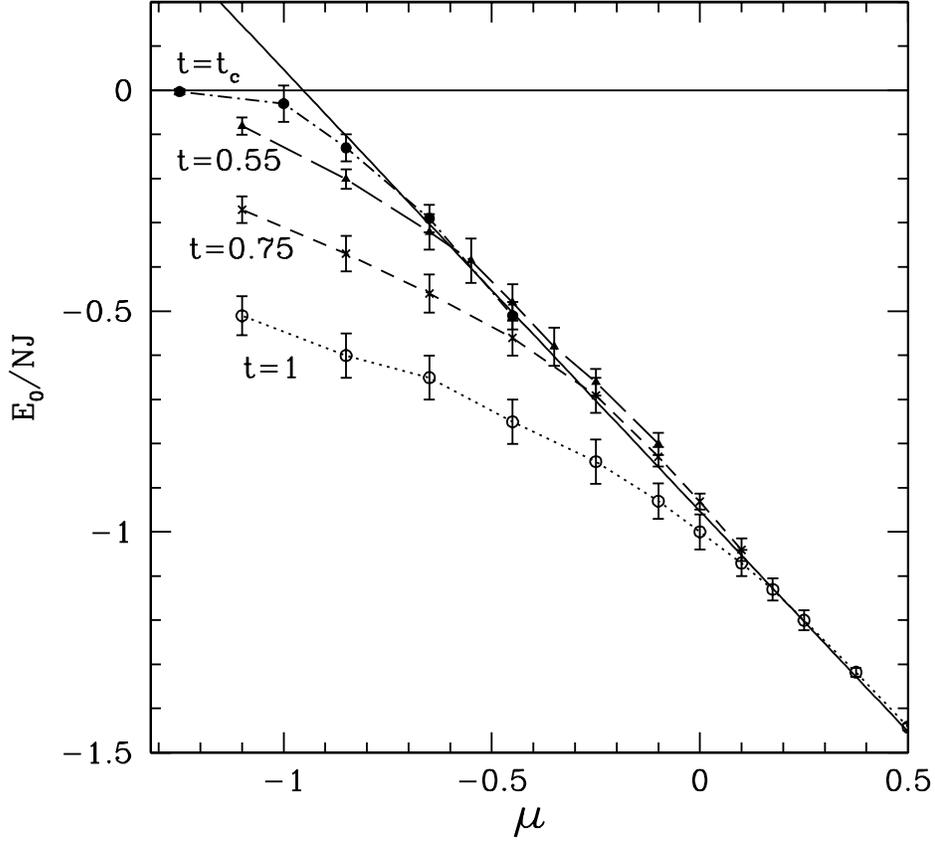,width=15cm}}}
\par
\caption{The ground state energy per site $E_0/NJ$
versus $\mu$ for $t=0.4638$, 0.55, 0.75 and 1. The solid line is the result at
half-filling\cite{zwh98}.
}
\label{fig_e0}
\end{figure}

 Figure \ref{fig_e0} shows curves of
$E_0$ versus $\mu$ for $t=0.46, 0.55, 0.75, 1.0$. The solid line in this
figure is the result for half-filling ($n=1$) where the ground
state energy per site, for $J=J_{\perp}$, is $E_0/NJ =
 -0.578043-3/8-\mu$.\cite{zwh98} 
 For $\mu \gtrsim $ 0.1-0.2 our ground state energy is very close to that for
 half-filling, for all $t$ considered. 
 
\begin{figure}[ht] 
\vspace{0cm}
\centerline{\hbox{\psfig{figure=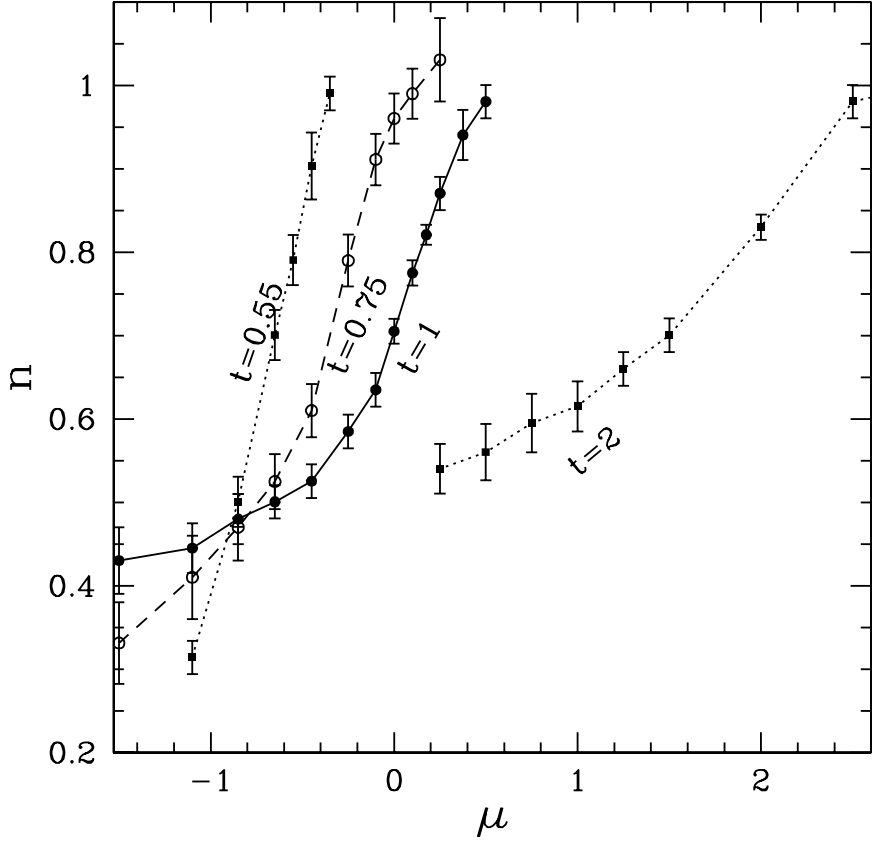,width=15cm}}}
\par
\caption{The electron density $n$ versus $\mu$ for $t=0.55$, 0.75, 1 and 2.
}
\label{fig_n}
\end{figure}

Figure \ref{fig_n} shows curves of electron density versus $\mu$
for $t=0.55, 0.75, 1, 2$ (setting $J=1$). 
For $\mu \gtrsim -0.7$ the electron density decreases with increasing
$t$, for constant $\mu$; or equivalently for fixed $n$ the value of
$\mu$ increases with $t$. Around $\mu\simeq -0.8$ there is a crossover region where
the value of $n$ is insensitive to $t$, and below this value of $\mu$
the dependence of $n$ on $t$ is the opposite to that discussed above.
As $t$ decreases the curves of $n$ versus $\mu$ steepen, and at the critical
point for phase separation, estimated to be $t_c=0.4638$\cite{rom00},
we expect a discontinuous drop away from $n=1$ to develop 
at some critical separation value
$\mu_c$. Our series data are not currently expressed as analytical functions
of $\mu$, and do not allow us to explore the phase separation phenomenon any further here.

\begin{figure}[ht] 
\vspace{0cm}
\centerline{\hbox{\psfig{figure=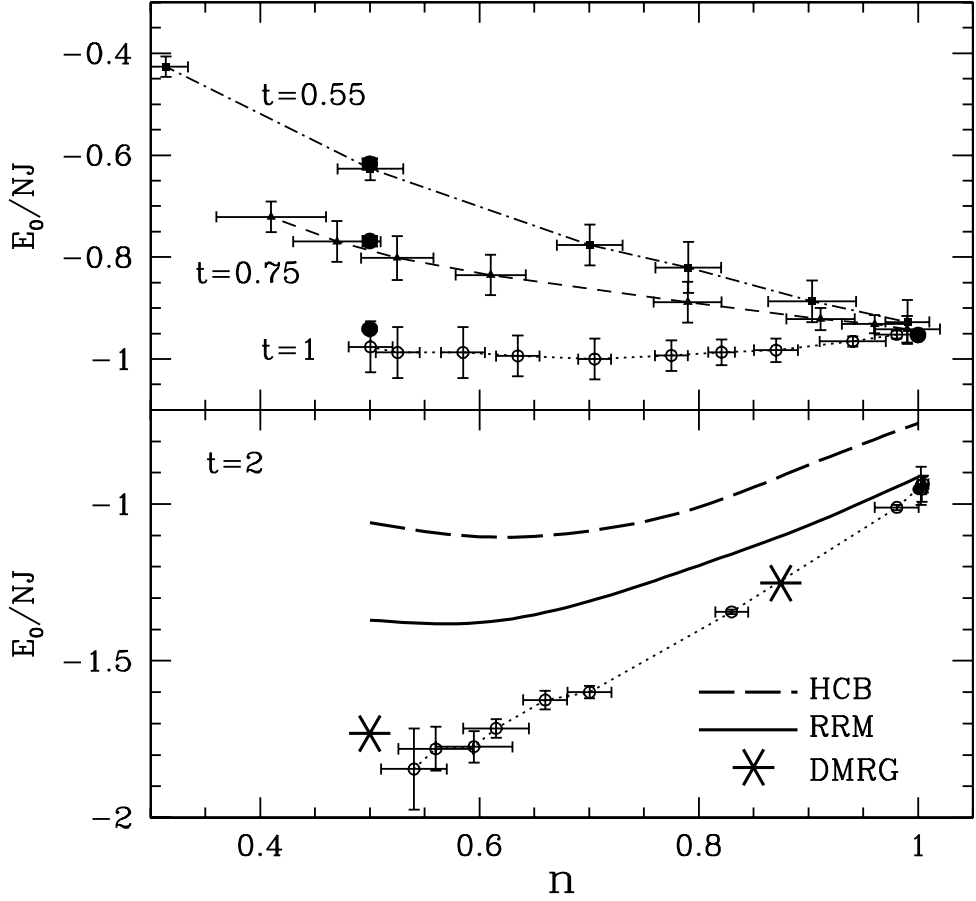,width=15cm}}}
\par
\caption{The ground state energy versus electron density $n$ for $t=0.55$ ,0.75, 1
(upper window) and 2 (lower window).
The large solid circles at quarter filling and half-filling 
are the results of series expansions at quarter filling\cite{oit99} (for $t=0.55$, 0.75 and 1)
and
series expansions for the Heisenberg ladder\cite{zwh98}. The bold long dashed line,
solid line and stars are the results of HCB, RRM and DMRG\cite{sie98} for $t=2$.
}
\label{fig_E0_n}
\end{figure}

We have used the data of Figure \ref{fig_n} to compute the ground state energy as a function of $n$, and
these results are shown in Figure \ref{fig_E0_n}. For ease of comparison with previous work
we have subtracted the chemical potential term $\mu n$ from the energy.
Figure \ref{fig_E0_n} also serves to compare our results with previous
work. The filled circles at $n=0.5$ are our previous results\cite{oit99}
for quarter-filling, for $t=0.55, 0.75, 1.0$. The dashed and solid lines
are results obtained from a hard-core boson approximation (HCB) and
a recurrence-relation method (RRM) for $t=2$ by Sierra {\it et al.}\cite{sie98} The crosses
are Density Matrix Renormalization Group results\cite{sie98}, also for
$t=2$. The latter agree very well with our series estimates, whereas
the analytic approximations clearly give too high an energy.

\begin{figure}[ht] 
\vspace{0cm}
\centerline{\hbox{\psfig{figure=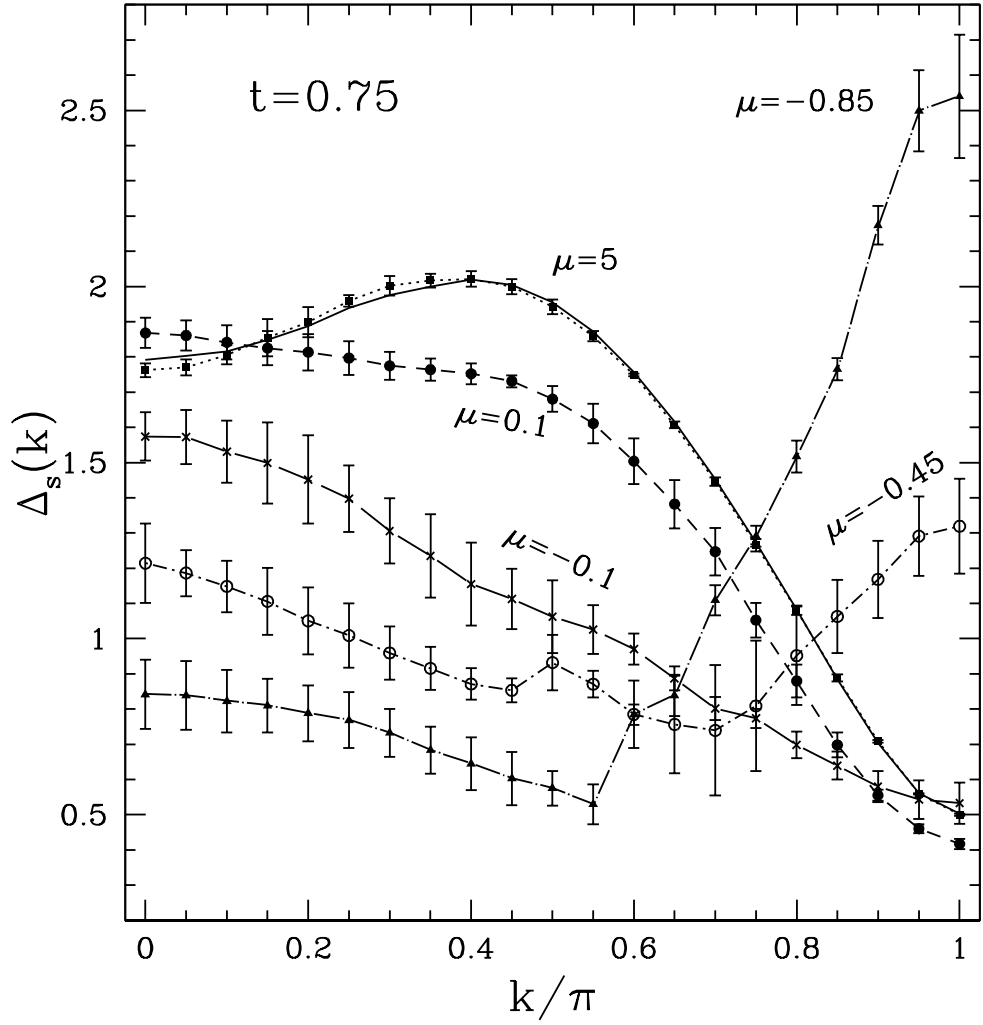,width=15cm}}}
\par
\caption{The triplet spin dispersion relation for $t=0.75$
and different $\mu$. Also presented are 
results for the Heisenberg spin ladder
for $J=J_{\perp}$ (solid line)\cite{zwh98}.
}
\label{fig_triplet_gap}
\end{figure}

Next we consider the dispersion relation
$\Delta_s (k)$ for triplet spin excitations.  Figure \ref{fig_triplet_gap} shows the triplet 
spin dispersion curve $\Delta_s (k)$
for $t=0.75$, and $\mu = 5$, 0.1, -0.1, -0.45, -0.85 which correspond to
$n=1.000(2)$, 0.99(3) 0.91(3) 0.61(3), 0.47(4), respectively. The solid curve is the
dispersion curve for the Heisenberg spin ladder with $J=J_{\perp}$\cite{zwh98}.
This is the dispersion curve for half-filling, and agrees very well with the present
results for $\mu=5$. The figure again shows that the triplet dispersion is very
sensitive to doping, with a significant change already for 1\%
doping ($\mu=0.1$). At small doping the minimum triplet energy remains
at $k=\pi$, as for half-filling, but for large doping it shifts to
intermediate $k$ values. The series become more irregular and the
error bars are correspondingly larger.
For $\mu=-0.85$, the shift in the minimum energy has become
a dramatic effect.

\begin{figure}[ht] 
\vspace{0cm}
\centerline{\hbox{\psfig{figure=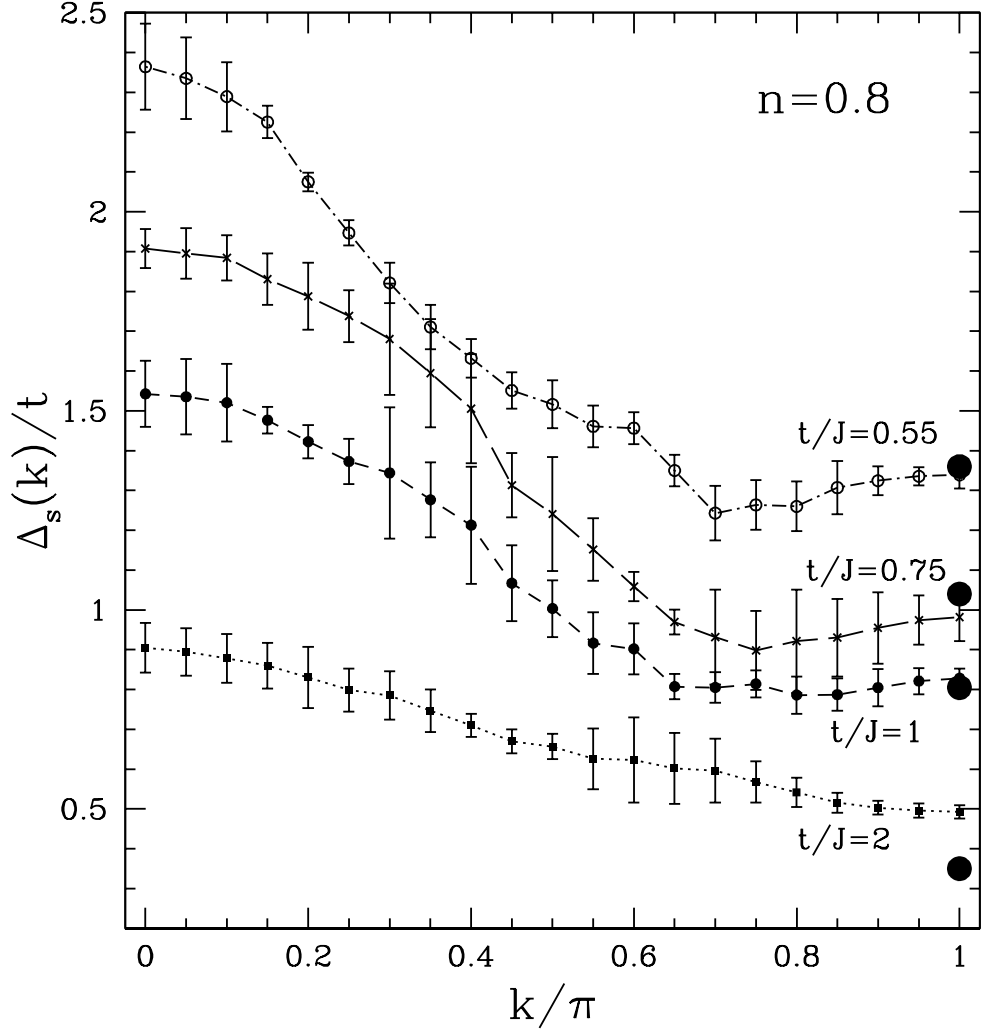,width=15cm}}}
\par
\caption{The triplet spin dispersion relation for $t/J=0.55$, 0.75, 1 and 2
and $n=0.8$. The full solid points are the finite lattice results for the
 minimum gap [Ref. \protect\onlinecite{hay96}].
}
\label{fig_triplet_gap_np8}
\end{figure}

Figure \ref{fig_triplet_gap_np8} shows the triplet 
spin dispersion curve $\Delta_s (k)$
for $t=0.55$, 0.75, 1 and 2, and $\mu$ is chosen to be -0.54, -0.24, 0.14, 1.9 respectively,  so that
$n\simeq 0.8$. We also show, in the Figure, previous results
for the minimum energy gap obtained from exact diagonalizations by 
Hayward {\it et al.}\cite{hay96} As can be
seen, the agreement is excellent.

\begin{figure}[ht] 
\vspace{0cm}
\centerline{\hbox{\psfig{figure=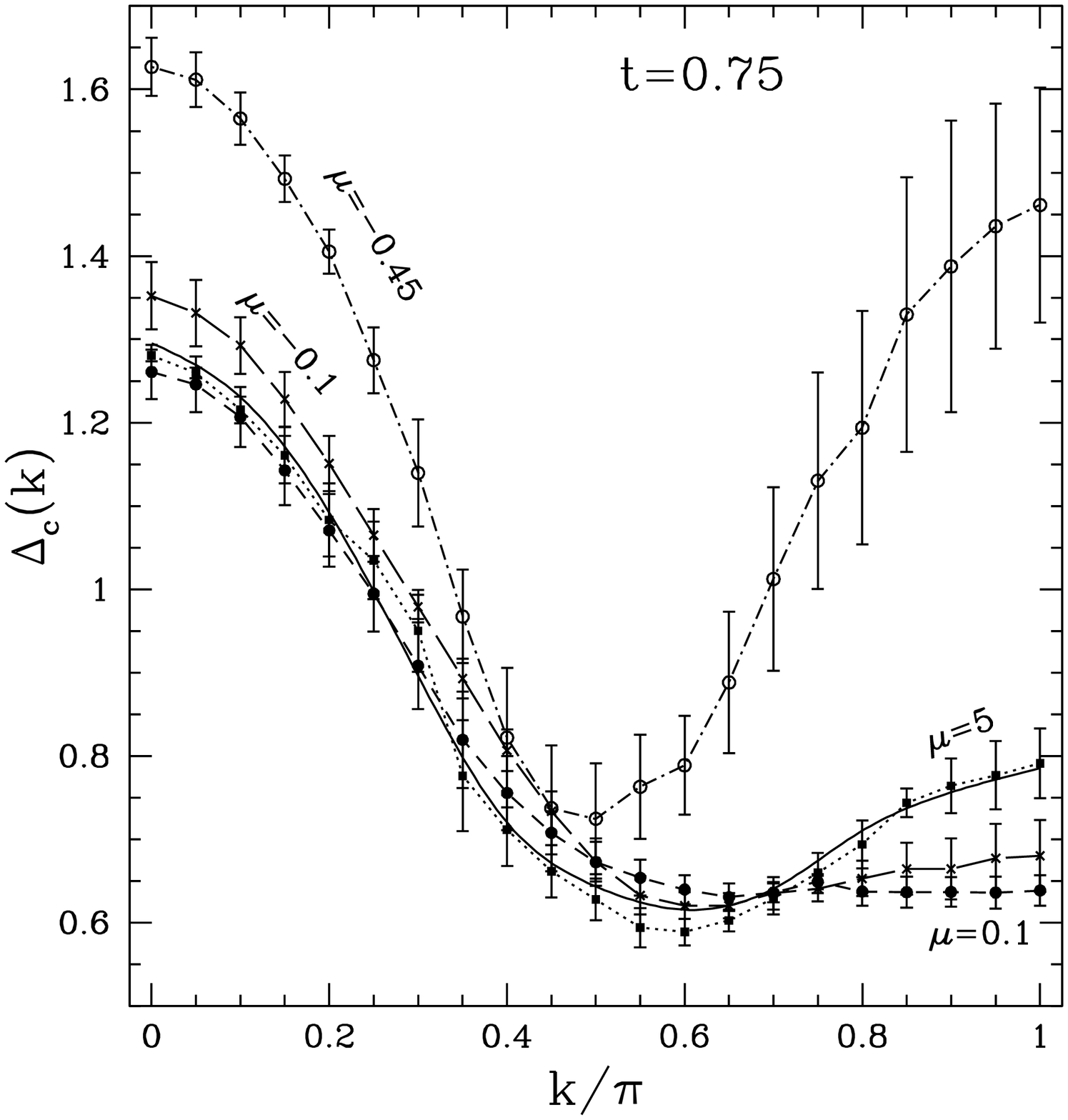,width=15cm}}}
\par
\caption{The dispersion relation for the one-hole bonding excitation at
$t=0.75$ and different $\mu$. Also presented are the
results at half-filling (solid line)\cite{cjh98}.
}
\label{fig_1h_bonding}
\end{figure}

Finally we turn to charge excitations, in particular the 1-hole bonding
excitation (which couples rung states 1 and 6, or 1 and 7 in Table 1).
These excitations correspond to creation or destruction of a hole in the ground
state. Figure \ref{fig_1h_bonding} show the dispersion $\Delta_c (k)$, again for
$t=0.75$ and various $\mu$, together with the corresponding
dispersion curve for the Heisenberg spin ladder (solid line)\cite{cjh98}.
This curve, which corresponds to half-filling, again agrees very well with our present results for
$\mu=5$. A small amount of doping (up to 10\%)
has little effect on the dispersion curve for $k\lesssim \pi/2$,  but results in a
noticeable flattening for larger $k$.
For $\mu =-0.45$ (corresponding to $n\simeq 0.6$) there is a pronounced
minimum at $k\simeq \pi/2$ and a large increase in excitation energy for
larger $k$. 
Troyer {\it et al.} have previously noted this substantial shift in the
minimum of the dispersion curve for the bonding and antibonding bands, and
have identified the minima with the Fermi points of the
respective quasiparticle bands.

\section{\label{sec:con}Discussion and Summary}

The method of linked-cluster perturbation expansions for strongly
interacting lattice electron models has, until now,
been restricted to systems with one electron per site (half-filling)
or to one or two holes in the half-filling state. In this paper we show,
for the first time, how this constraint can be relaxed, and we demonstrate
its reliability in the case of the 2-leg $t-J$ ladder, where comparisons
with other methods can be made. The agreement is found to be well 
within the error estimates.

We present a number of new results for the $t-J$ ladder, for the ground
state energy and electron density as functions of chemical potential, and for the
ground state energy and some excitation energies for different values
of electron density. Results are presented for a case in which the rung 
interactions are stronger than the chain interactions (by a factor of 4)
and also for the isotropic case.

This work opens up new possibilities for exploring other aspects of strongly correlated
electron models with a finite hole density, and we intend to take up
some of these in future work.
The series approach is not subject to finite-size effects, unlike exact
diagonalization, the DMRG, or (to a lesser extent) Monte Carlo simulations;
and it provides a versatile complementary method to these other numerical techniques.
It also suffers from some limitations, of course, for instance, it does
not allow one to explore the phase separated regime, where the electron density
dependence on the chemical potential is discontinuous.

\begin{acknowledgments}
The work of Z.W.,  C.J.H. and J.O. forms part of 
a research project supported by a grant
from the Australian Research Council, R.R.P.S. is
supported in part by NSF grant number DMR-9986948. 
We have benefited from discussions with
Professor O.P. Sushkov.
The computation has been performed on the AlphaServer SC
 computer. We are grateful for the computing resources provided
 by the Australian Partnership for Advanced Computing (APAC)
National Facility.
\end{acknowledgments}


\newpage
\bibliography{basename of .bib file}


\begin{table}
\caption{The nine rung states and their energies, where $\kappa = \sqrt{4 h^2 + ( J_{\perp} + 2 \mu)^2 }$.
}\label{tab1}
\begin{tabular}{|c|c|c|c|}
\hline\hline
\multicolumn{1}{|c|}{No.} &\multicolumn{1}{c|}{Eigenstate}
&\multicolumn{1}{c|}{Eigenvalue} &\multicolumn{1}{c|}{Name} \\
\hline
1  &  $(\kappa-J_{\perp} -2 \mu) \mid 00 \rangle -
  \sqrt{2} h (\mid \uparrow \downarrow \rangle - \mid \downarrow \uparrow \rangle )$ &
  $-(J_{\perp}+2\mu + \kappa)/2$  &  lower energy singlet  \\
\hline
2  & $\mid \downarrow \downarrow \rangle  $  &    $-2\mu$  &  triplet ($S^z_{\rm tot}=-1$)  \\
\hline
3  & $\frac{1}{\sqrt 2}(\mid \uparrow \downarrow \rangle + \mid \downarrow \uparrow \rangle )$ & $-2\mu$  &  triplet ($S^z_{\rm tot}=0$)  \\
\hline
4  & $\mid \uparrow \uparrow \rangle  $  &    $-2\mu$  &  triplet ($S^z_{\rm tot}=1$)  \\
\hline
5  & $-(J_{\perp}+\kappa+2 \mu) \mid 00 \rangle -
  \sqrt{2} h (\mid \uparrow \downarrow \rangle - \mid \downarrow \uparrow \rangle )$  &
  $(\kappa-J_{\perp}-2\mu)/2$  &  higher energy singlet  \\
\hline
6  & $\frac{1}{\sqrt 2}(\mid 0 \downarrow \rangle + \mid \downarrow 0 \rangle )$ & $-t_{\perp}-\mu$  & electron-hole bonding ($S^z_{\rm tot}=-\frac{1}{2}$)  \\
\hline
7  & $\frac{1}{\sqrt 2}(\mid 0 \uparrow \rangle + \mid \uparrow 0 \rangle )$ & $-t_{\perp}-\mu$  & electron-hole bonding ($S^z_{\rm tot}=\frac{1}{2}$)  \\
\hline
8  & $\frac{1}{\sqrt 2}(\mid 0 \downarrow \rangle - \mid \downarrow 0 \rangle )$ & $t_{\perp}-\mu$  & electron-hole antibonding ($S^z_{\rm tot}=-\frac{1}{2}$)  \\
\hline
9  & $\frac{1}{\sqrt 2}(\mid 0 \uparrow \rangle - \mid \uparrow 0 \rangle )$ & $t_{\perp}-\mu$  & electron-hole antibonding ($S^z_{\rm tot}=\frac{1}{2}$)  \\
\hline\hline
\end{tabular}
\end{table}

\end{document}